\documentclass[twocolumn,aps,amsfonts,prl]{revtex4}
\usepackage{graphicx}
\usepackage{mathrsfs}
\usepackage{amsmath}
\usepackage{mathrsfs}
\usepackage{amssymb}
\usepackage{color}
\usepackage[nice]{nicefrac}
\usepackage{float}
\begin{document}

\def\be{\begin{eqnarray}}
\def\ee{\end{eqnarray}}
\def\nn{\nonumber}
\def\qnum#1#2{\lfloor#1\rfloor_{#2}}

\title{Topological Quantum Computing with Read-Rezayi States}

\author{L. Hormozi$^1$}
\author{N.E. Bonesteel$^2$}
\author{S.H. Simon$^3$}
\affiliation{
$^1$Joint Quantum Institute, National Institute of Standards and Technology and University of Maryland, Gaithersburg, MD 20899\\
$^2$Department of Physics and NHMFL, Florida State University, Tallahassee, FL 32310 \\
$^3$Rudolf Peierls Centre for Theoretical Physics, Oxford University, 1 Keble Road, Oxford OX1 3NP, UK}

\begin{abstract}
Read-Rezayi fractional quantum Hall states are among the prime
candidates for realizing non-Abelian anyons which in principle can
be used for topological quantum computation. We present a
prescription for efficiently finding braids which can be used to
carry out a universal set of quantum gates on encoded qubits based
on anyons of the Read-Rezayi states with $k>2$, $k\neq4$. This
work extends previous results which only applied to the case $k =
3$ (Fibonacci) and clarifies why in that case gate constructions
are simpler than for a generic Read-Rezayi state.
\end{abstract}

\pacs{PACS numbers: }

\maketitle

Non-Abelian anyons \cite{moore91} --- quasiparticle excitations
obeying so-called non-Abelian statistics --- are conjectured to
emerge in certain two-dimensional quantum systems. In these
systems, when well-separated anyons are present, there is a ground
state degeneracy that grows exponentially with the number of
anyons. Furthermore, if these anyons remain well-separated,
different states in the ground state manifold cannot be
distinguished by local measurements --- thus rendering this space
immune from decoherence due to any local perturbations.

When non-Abelian anyons are exchanged, the process is described by
a multidimensional unitary operation (instead of a single phase)
acting on the degenerate space. Certain unitary operations can
then be carried out by dragging anyons around one another,
``braiding" their worldlines in 2+1 dimensional space-time. As
long as the anyons are kept sufficiently far apart during this
process, the resulting unitary operation will be identical for any
two topologically equivalent braids.

Recent interest in non-Abelian anyons has focused on the
possibility of using them for topological quantum computation ---
a form of quantum computation which exploits the topological
robustness of braiding and the protection of the degenerate
Hilbert space against decoherence to process and store quantum
information in an intrinsically fault-tolerant
way~\cite{kitaev,freedman,nayak08}.  This paper is concerned with
the problem of finding specific braiding patterns which can be
used to carry out universal quantum computation for a class of
non-Abelian anyons described by $\mathfrak{su}(2)_k$
Chern-Simons-Witten theories.

The theory of $\mathfrak{su}(2)_k$ anyons provides the
mathematical description \cite{slingerland01ardonne06} (up to
Abelian phases not relevant here) of the braiding properties of
quasiparticle excitations in the Read-Rezayi \cite{read99rezayi06}
sequence of fractional quantum Hall~(FQH) states.  Here the
parameter $k$ (the ``level") is a positive integer characterizing
the state. For example, the $k = 2$ state is the Moore-Read state
\cite{moore91}, believed to describe the FQH plateau observed at
filling fraction $\nu = 5/2$ \cite{nayak08}, and the $k=3$ state
may describe the FQH plateau observed at $\nu = 12/5$
\cite{read99rezayi06}.  Bosonic Read-Rezayi states may also be
realizable in rotating Bose gases~\cite{cooper01}, and model spin
Hamiltonians have been constructed~\cite{fendley05levin05} for
which the low-energy quasiparticle excitations can be described by
any consistent achiral theory of non-Abelian anyons, including
(doubled) $\mathfrak{su}(2)_k$, suggesting the possibility of
realizing $\mathfrak{su}(2)_k$ anyons in exotic spin liquids.

It has been shown that $\mathfrak{su}(2)_k$ anyons with $k = 3$ or
$k > 4$ can be used to carry out universal quantum computation
just by braiding anyons~\cite{freedman}. In previous
work~\cite{us,xu08} a number of different prescriptions have been
given for explicitly constructing braids to carry out universal
quantum computation using $\mathfrak{su}(2)_3$ anyons --- anyons
which are, for our purposes, essentially equivalent to the
so-called Fibonacci anyons~\cite{nayak08}. In the present work, we
extend these results to all $\mathfrak{su}(2)_k$ anyons with
$k\geq3$, $k\neq4$.

$\mathfrak{su}(2)_k$ anyons carry a quantum number resembling
ordinary spin referred to here as (topological)
``charge"~\cite{nayak08,qgroups}. For the level $k$ theory the
allowed values of this charge include all integers and half
integers between 0 and $k/2$. Similar to ordinary spin, there are
rules for combining topological charge which specify the possible
total charge of objects formed when two or more anyons are
combined. For $\mathfrak{su}(2)_k$ anyons the fusion rule is a
truncated version of the usual triangle rule for adding angular
momenta,
\be\label{fusion}
s_1 \otimes s_2 &=& |s_1 - s_2| \oplus (|s_1 - s_2| +
1)\oplus\cdots\\\nn &\oplus& \min\left(s_1 + s_2, k -
(s_1+s_2)\right).
\ee

From this fusion rule it can be shown that, asymptotically, the
dimensionality of the Hilbert space of $N$ identical anyons grows
as $d_k^N$ where $d_k$ is known as the quantum dimension of the
particles \cite{nayak08,qgroups}. For charge 1/2 anyons $d_k =
\lfloor2\rfloor_q$, where the $q$-integer $\lfloor m\rfloor_q$ is
defined as $\lfloor m\rfloor_q =
(q^{m/2}-q^{-m/2})/(q^{1/2}-q^{-1/2})$ and $q = e^{i2\pi/(k+2)}$.
For example, in the case of ordinary spin 1/2 particles
(corresponding to $\mathfrak{su}(2)_k$ anyons when $k\to\infty$)
the quantum dimension is 2, as expected, while for Fibonacci
anyons ($k = 3$), where the dimensionality of the Hilbert space
grows as the Fibonacci sequence, the quantum dimension is the
golden ratio.

\begin{figure}[t]
\begin{center}
\includegraphics[width = 3.5in]{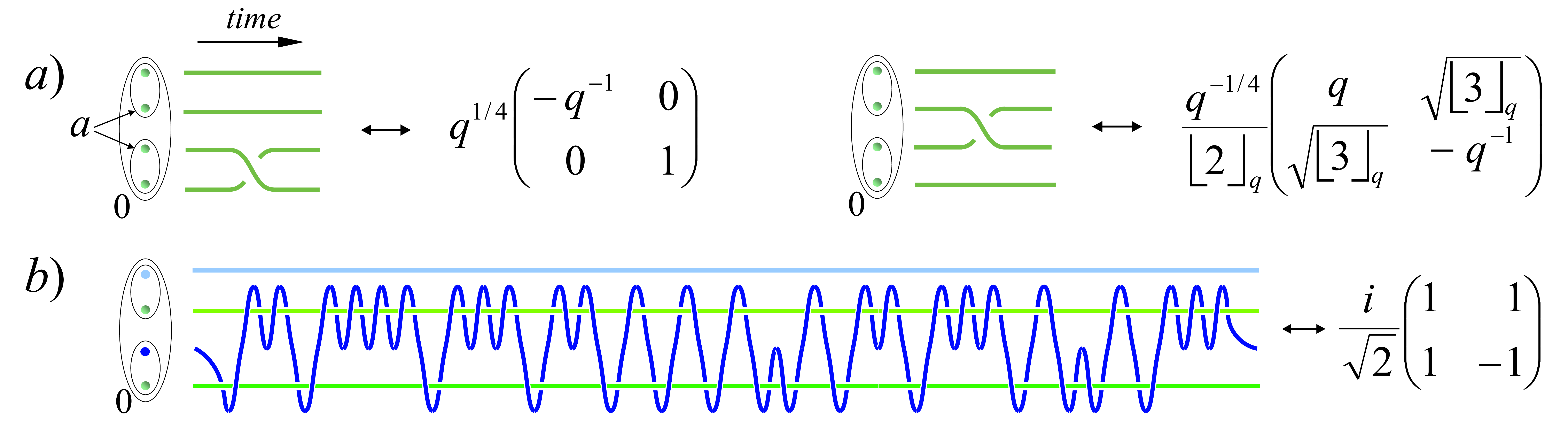}
\caption{(a) Encoded qubits and elementary braid matrices for
$\mathfrak{su}(2)_k$ anyons, and (b) a sample single-qubit gate.
Here, and in subsequent figures, groups of anyons in charge
eigenstates are enclosed in ovals labeled by the charge. Qubits
are encoded in the two-dimensional Hilbert space of four charge
$1/2$ anyons with total charge 0. The matrices shown, in the basis
labeled by $a = \{0,1\}$, correspond to the elementary braid
operations, also shown. Within the encoded qubit space, braiding
the top two anyons produces the same unitary operation as braiding
the bottom two. The braid depicted in (b) consists of 80
interchanges and approximates a Hadamard gate for
$\mathfrak{su}(2)_5$ anyons with an accuracy of~$\sim~2.2\times10^{-6}$ (measured using operator norm, see \cite{us}).}
\label{qubits}
\end{center}
\end{figure}

For all finite $k \ge 2$, the quantum dimension is an irrational
number between 1 and 2.  It follows that the Hilbert space of $N$
charge 1/2, $\mathfrak{su}(2)_k$ anyons cannot be decomposed into
a tensor product of smaller subsystems. To carry out quantum
computation within the standard ``qubit plus quantum gate"
framework one must therefore encode qubits using several anyons.
Here we encode qubits in the two-dimensional Hilbert space of four
charge 1/2 anyons with total charge
0~\cite{freedman03}. Referring to Fig.~\ref{qubits},
we choose the logical $|0\rangle$ and $|1\rangle$ for this qubit
to be states for which the total charge of the two bottommost (or,
equivalently, the two topmost) anyons is 0 and 1, respectively.
Note that the fusion rule (\ref{fusion}) indicates that when the
constraint that the total charge be 0 is relaxed the Hilbert space
of four charge 1/2 anyons is five-dimensional for $k = 3$ and
six-dimensional for $k>3$. The states that do not correspond to
qubit states, i.e. states with total charge 1 or 2, are then {\it
non-computational states}. The existence of these states implies
the possibility of making transitions from the qubit space to the
non-computational space, i.e. leakage errors, which pose a major
problem in constructing quantum gates.

To carry out universal quantum computation it must be possible to
perform arbitrary single-qubit gates as well as at least one
entangling two-qubit gate. It has been shown that for
$\mathfrak{su}(2)_k$ anyons when $k\geq 3$, $k\neq4, 8$, any
desired single-qubit gate can be approximated, to any desired
accuracy, by weaving a charge 1/2 anyon around two others within a
qubit~\cite{freedman,simon06}. In practice, braids which approximate a
given single-qubit gate can be found by carrying out a search over
braids up to a given length and choosing those which produce
unitary operations closest to the desired target gate.  As an
example, Fig.~\ref{qubits} shows a braid which is the result of a bidirectional search~\cite{zikos08} and which approximates a
Hadamard gate using $\mathfrak{su}(2)_5$ anyons to an accuracy of
1 part in 10$^6$. Given the ability to perform a search this deep, further accuracy can
always be systematically achieved by applying the Solovay-Kitaev
algorithm~\cite{us}.  Note that for single-qubit gates there is no
danger of leakage errors because braiding within a qubit cannot
change its total topological charge.

When searching for braids we are performing a discrete search over
a continuous space.  The dimensionality of the space being
searched, $D$, is the main factor which determines the efficacy of
these searches. If the number of distinct braids one can search is
$N_b$, the typical error for approximating a given
target gate will be $\sim N_b^{-1/D}$. Thus, for fixed $N_b$,
increasing $D$ greatly reduces the accuracy of the braids one can
obtain.  For single-qubit gates the search space is $SU(2)$ with
$D= 3$ which is sufficiently small to allow us to find highly
accurate braids, as demonstrated by the above example.

Two-qubit gates are significantly harder to construct, both due to
the possibility of leakage errors, and the fact that for eight
anyons the search space is $SU(13)$ for $k=3$ and $SU(14)$ for
$k>3$ (with dimensionalities of 168 and 195, respectively). The
key idea for efficiently finding braids for two-qubit gates,
introduced in~\cite{us} and common to more recent
approaches~\cite{xu08}, is to reduce the two-qubit gate problem
to one or more effective ``single-qubit" problems, i.e. problems
in which one is searching over braids which involve only three
objects at time (where an object can be either a single anyon, or
a collection of anyons braided as a single entity), and for which
the search space is $SU(2)$.  In this case, as emphasized above,
excellent approximations for any desired target operation can be
obtained.

In addition to reducing all searches to $SU(2)$, previous work has
focused on the case $k = 3$.  To see why $k = 3$ is special note
that in this case, for anyons with topological charge $1$, the
only non-trivial fusion rule is $1 \otimes 1 = 0 \oplus 1$, which
is the fusion rule of the Fibonacci anyons~\cite{nayak08}.  All
previous gate constructions have exploited in one way or another
the unique feature of Fibonacci anyons that there is only one
nontrivial value of the topological charge. Thus any collection of
Fibonacci anyons either has topological charge 0 and is therefore
``neutral" (it does not induce any non-Abelian transitions if
braided as a single cohesive object) or has topological charge 1,
in which case it behaves as a single Fibonacci anyon.

\begin{figure}[t]
\begin{center}
\includegraphics[width = 3.5in]{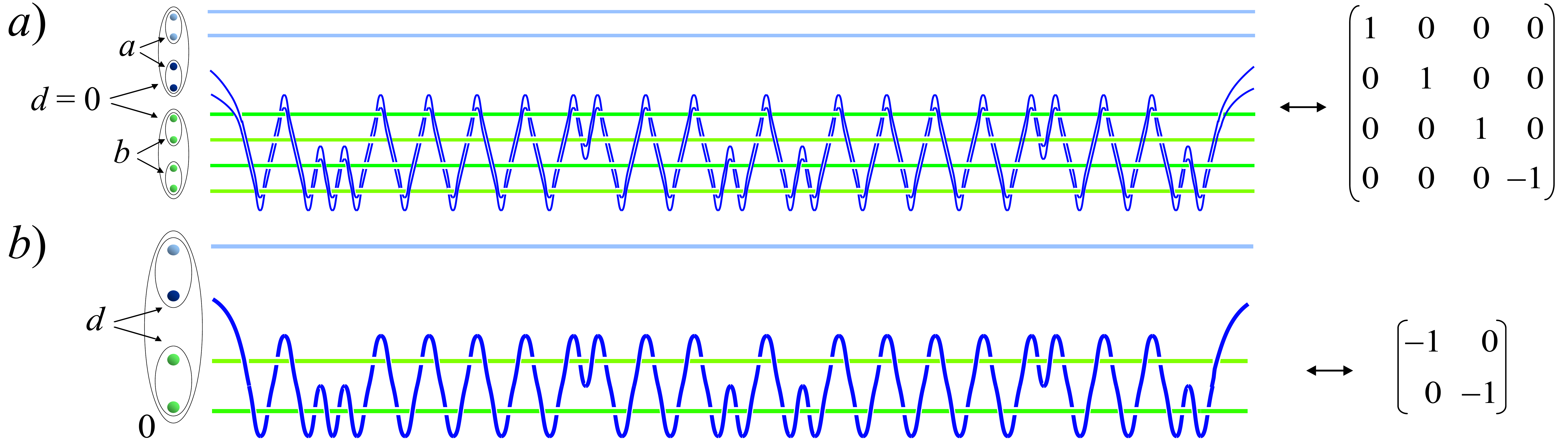}
\caption{(Color Online). ``Effective qubit" gate construction for
$\mathfrak{su}(2)_3$ anyons.  Part (a) shows a braid in which a
pair of anyons from the control qubit (shown in blue) weaves
around pairs of anyons in the target qubit (shown in green). When
either the control or target qubits are in the state $|0\rangle$
this braid produces the identity operation. When both control and
target qubits are in the state $|1\rangle$ the braid consists of
weaving a charge 1 anyon around two other charge 1 anyons, as
shown in (b). For $\mathfrak{su}(2)_3$ anyons the braiding
properties of charge $1/2$ anyons and charge 1 anyons are the same
(up to irrelevant Abelian phases), and the problem is reduced to
that of finding a particular single-qubit gate by carrying out a
search in $SU(2)$. The braid shown in (b) approximates a negative
identity matrix, in the basis labeled by $d = \{0,1\}$, for
$\mathfrak{su}(2)_3$ anyons with an accuracy
of~$\sim1.6\times10^{-5}$. As a consequence, the full braid shown in (a) approximates a controlled-$Z$ gate, shown in the basis labeled by $ab =
\{00,01,10,11\}$, with an accuracy of~$\sim1.4\times 10^{-5}$.}
\label{bigqubit}
\end{center}
\end{figure}

We illustrate the usefulness of these features with a simple
two-qubit gate construction \cite{thesis} (see also
\cite{xu08}). Figure~\ref{bigqubit}(a) shows a two-qubit braid
in which a pair of anyons from the control qubit (the control
pair) is woven around \emph{pairs} of anyons in the target qubit,
before returning to its original position. When the control qubit
is in the state $|0\rangle$ the control pair has charge 0 and,
because weaving a charge 0 object around other anyons does not
induce any non-Abelian transitions, the result of this operation
will be trivial (i.e. the identity). Similarly, when the target
qubit is in the state $|0\rangle$, the control pair is woven
around objects with total charge 0 and the result will again be
the identity (regardless of the state of the control qubit). Thus,
by construction this weave acts as the identity on the two-qubit
states $|0\rangle|0\rangle$, $|0\rangle|1\rangle$, and
$|1\rangle|0\rangle$, with the only nontrivial case being
$|1\rangle|1\rangle$ (here the first qubit is the control qubit).
To construct an entangling two-qubit gate it is then necessary to
find a particular weave of the form shown in
Fig.~\ref{bigqubit}(a) which returns the state
$|1\rangle|1\rangle$ to itself while acquiring a nontrivial phase
with respect to the state $|1\rangle|0\rangle$, thus producing a
controlled rotation of the target qubit.

Finding such a weave is straightforward for Fibonacci anyons
($k=3$).  In this case the fusion rule implies that the Hilbert
space of four charge 1 objects with total charge 0 is
two-dimensional and the problem reduces to that of searching for a
particular single-qubit operation acting on the ``effective qubit"
shown in Fig.~\ref{bigqubit}(b). The states of this ``qubit" are
determined by the label $d$ which can be either 0 or 1. To ensure
there are no transitions between the encoded qubit states ($d =
0$) and the non-computational states ($d = 1$), the resulting
unitary operation must be diagonal in $d$. As an example,
Fig.~\ref{bigqubit}(b) shows a braid which approximates a negative
identity matrix. If one follows this braiding pattern by weaving
the control pair around pairs of anyons in the target qubit (as
shown in Fig.~\ref{bigqubit}(a)), the $|1\rangle|1\rangle$ state
acquires a phase of $-1$ and the resulting two-qubit gate is a
controlled-$Z$ gate which is equivalent to a CNOT, up to
single-qubit rotations.

We now turn to the case $k>3$.  In this case the fusion rule for
combining two charge 1 objects is $1\otimes1 = 0 \oplus 1 \oplus
2$. This implies that $d$, the overall charge of the original
qubits shown in Fig.~\ref{bigqubit}, can now take three different
values. Hence, when $k>3$ the unitary operations corresponding to
braids of the form shown in Fig.~\ref{bigqubit}(a) are elements of
$SU(3)$, not $SU(2)$. It is in principle possible to carry out
a search in $SU(3)$, but because the dimensionality of
the search space is 8, rather than 3, it is significantly less
efficient than searching $SU(2)$.

\begin{figure*}[t]
\begin{center}
\includegraphics[width=7in]{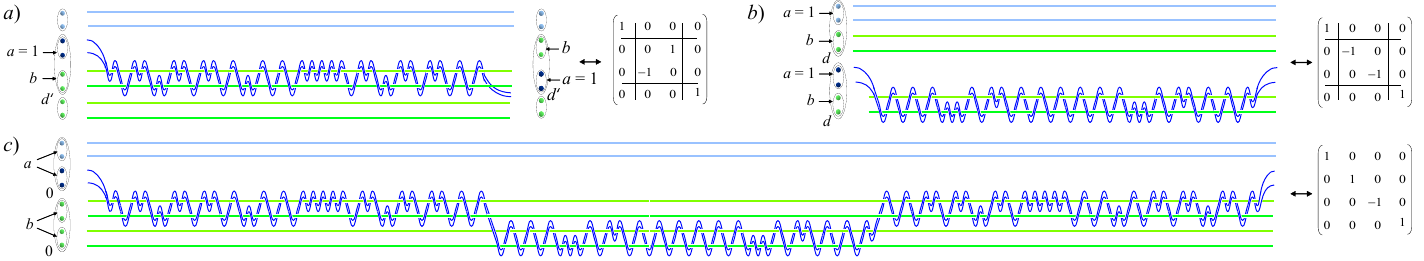}
\caption{(Color Online) Controlled-Phase gate construction for
$\mathfrak{su}(2)_5$ anyons. Part (a) shows a swap braid which
effectively exchanges the control pair (shown in blue) with the
topmost pair of anyons in the target qubit (green). This braid
approximates the unitary operation shown in the figure with an
accuracy of~$\sim 5.1\times10^{-6}$. Note that the matrix shown is
expressed using inequivalent left and right bases states labeled
$bd' = \{10,01,11,12\}$ at both the start and end of the braid.
Part (b) shows a braid which acts on the intermediate state
produced by the swap braid and gives the state with $a=1$ $b=0$ a
phase of $-1$. The corresponding unitary operation shown in the
figure in the basis labeled by $bd = \{10,01,11,12\}$ is
approximated with an accuracy of~$\sim 7.0\times10^{-6}$. (c) Full
controlled-$(-Z)$ gate, in the basis labeled by $ab =
\{00,01,10,11\}$, with an accuracy of~$\sim 7.4\times10^{-7}$.}
\label{kcnot5}
\end{center}
\end{figure*}

In general we find that for $k>3$ it is impossible to construct a
leakage free two-qubit entangling gate by performing a single
braid in which three objects are braided and the search space is
$SU(2)$~\cite{footnote}.  However, it {\it
is} possible to construct such a gate by breaking the construction
into three steps, as illustrated in Fig.~\ref{kcnot5}.  In each of
these steps a pair of anyons from the control qubit (again, the
control pair) is woven around two anyons in the target qubit.

Before describing the details of the construction we establish the
key fact that finding braids for each step only requires a search
in $SU(2)$. As before we need only consider the case when the
control pair has charge 1 ($a = 1$).  Each step then involves
weaving a charge 1 object around two charge 1/2 anyons. According
to the fusion rule (\ref{fusion}) the Hilbert space of these three
objects decouples into two one-dimensional sectors (with total
charge 0 and 2) and a single two-dimensional sector (with total
charge 1). The action on the one-dimensional sectors is determined
entirely by the winding number of the braid --- in particular if
we fix the winding number to be 0 the action is trivially the
identity~\cite{us}.  The only nontrivial action is then on the
two-dimensional sector, for which the relevant search space is
$SU(2)$.  For $\mathfrak{su}(2)_k$ anyons it is straightforward to
determine the relevant braid matrices~\cite{qgroups,thesis} and we find
that for all $k\ge 3$, $k\ne 4$ these matrices generate dense
covers of $SU(2)$, (this includes the case $k=8$ for which
braiding three charge 1/2 objects is {\it not} dense in $SU(2)$
\cite{freedman}). We therefore expect, and do indeed find, that
for each step in our construction carrying out a search can
produce braids which approximate the desired operations to an
accuracy of few parts in 10$^6$.

Now we turn to the actual construction. The first step consists of
a braid which effectively ``swaps" the control pair with a pair of
anyons in the target qubit. Figure~\ref{kcnot5}(a) shows this step
for $\mathfrak{su}(2)_5$ anyons. Referring to this figure, the
control pair starts from the bottom position in the control qubit,
weaves around the two topmost anyons in the target qubit, in the
end swapping positions with them. The specific braid shown is the
result of a search for a braid which generates a
unitary operation approximating the matrix shown in the figure.
(In this matrix --- and the one shown for step two --- the upper
left and lower right elements correspond to the one-dimensional
sectors with total charge 0 and 2, and the middle
$2\times 2$ block acts on the two-dimensional sector with total
charge 1, described above). This operation is
designed to effectively swap the control pair with the topmost
pair of anyons in the target qubit (labeled $b$), and when $a=b=1$
it does so without disturbing the quantum numbers of the system.
This means that if the initial state of the qubit is
$|1\rangle|1\rangle$ the final state will be the same but with the
control pair now swapped into the target.

The net effect of this swap operation is to take the system to the
intermediate state shown as the starting state in
Fig.~\ref{kcnot5}(b). In this intermediate state all the
information of the initial state of the two qubits is encoded in
the bottom four anyons.  In particular (assuming $a=1$) the total
charge of these anyons, labeled $d$ in the figure, must be $0$ if
$b=1$ (since in this case the swap operation is essentially the
identity) and 1 if $b=0$ (due to the fusion rules).

To perform an entangling two-qubit gate we need only induce a
phase shift between the $b=0$, and $b=1$ ($d = 0$) states when
$a=1$.  This is done in the second step of our construction, in
which the control pair is woven around the bottom two anyons and
returned to its starting position (see Fig.~\ref{kcnot5}(b)).  The
weave shown is the result of a search which produces
a unitary operation that is diagonal in $b$ and gives the state
$a=1$, $b=0$ a nontrivial phase ($-1$ for the braid shown).

In the third and final step the control pair is returned to its
original position in the control qubit by applying the inverse of
the swap braid. Putting all three steps together, the resulting
full braid is shown in Fig.~\ref{kcnot5}(c).  If the control qubit
is in the state $|0\rangle$, the control pair has charge 0 and the
effect of this braid is simply the identity operation.  If the
control qubit is in the state $|1\rangle$, this braid first swaps
the control pair into the target, then, if the target qubit was
initially in the state $|0\rangle$, gives the resulting
intermediate state a phase $e^{i\phi}$, and finally returns the
control pair to the control qubit. The full braid then
approximates a controlled-phase gate for which, if the control
qubit is in the state $|1\rangle$, the target qubit is rotated
about the $\hat{z}$ axis by the angle $\phi$. When $\phi = \pi$,
as is the case for the braid shown in the figure, this gate is a
controlled-$(-Z)$ gate which is equivalent to a CNOT, up to
single-qubit rotations.

We conclude by pointing out that for certain values of $k$ it is
possible to carry out step two of our construction with a finite
braid in such a way that the phase difference between the state
$a=1$, $b=0$ and $a=1$, $b=1$ ($d = 0$) is {\it exactly} $-1$.
Specifically, for $k = 8n-2$ this can be done by weaving the
control pair completely around the two bottom anyons in the target
qubit $n$ times.

We gratefully acknowledge Wayne Witzel for developing some of the
codes used to find the braids shown in this paper.  We also thank
the Aspen Center for Physics for its hospitality during the
completion of part of this work, and ICAM for providing travel
support (LH). LH is supported by the NIST/NRC postdoctoral program and NEB is supported by US DOE Grant No.
DE-FG02-97ER45639.


\begin{references}
\bibitem{footnote} One might consider constructing a two-qubit gate
by weaving the control pair around two anyons in the target qubit,
or weaving it around a pair and a single anyon. It can be shown
that the former cannot lead to an entangling two-qubit gate when
$k>3$~\cite{us} while the latter requires a search in $SU(2)\oplus
SU(2)$ with a six-dimensional search space.
\end{references}
\end{document}